\documentclass{article}
\usepackage{glas}
\usepackage{makeidx,subfigure}
\usepackage{graphicx}
\graphicspath{{eps/}{pics/}{figs/}}
\begin{document}
\begin{titlepage}{GLAS-PPE/97--03}{27 August 1997}

\title{Characterisation of low pressure VPE GaAs diodes before and after 24~GeV/c proton irradiation}

\author{R.L.~Bates\InstAnotref{glas}{\dagger}
C.~Da'Via\Instref{glas}
V.~O'Shea\Instref{glas}
C.~Raine\Instref{glas}
K.M.~Smith\Instref{glas}
R.Adams\Instref{epi}}
\Instfoot{glas}{Dept. of Physics \& Astronomy, University of Glasgow, UK}
\Instfoot{epi}{Epitronics Corporation, Phoenix, Az., USA}
\Anotfoot{\dagger}{Partially supported by a CASE award from the Rutherford Appleton Lab , UK}

\collaboration{On Behalf of the RD8 Collaboration}

\begin{abstract}

GaAs Schottky diode particle detectors have been fabricated upon Low Pressure Vapour Phase Epitaxial GaAs. The devices were characterised with both 
electrical and charge collection techniques. The height of the Ti-GaAs barrier used was determined via two electrical methods to be $0.81\pm0.005$ and $0.85\pm0.01$~eV. The current density was greater than that expected for an ideal Schottky barrier and the excess current was attributed to generation current in the bulk of the material. A space charge density of $2.8\pm0.2\times10^{14}$~cm$^{-3}$ was determined from capacitance voltage characterisation. The charge collection efficiency was determined from front alpha illumination and 60~keV gamma irradiation to be greater than 95\% at a reverse bias of 50~V. 

The diodes were characterised after an exposure to a radiation fluence of $1.25\times10^{14}$ 24GeV/c protons~cm$^{-2}$. The reverse current measured at 20$^{o}$~C increased from 90~nA to 1500~nA at an applied reverse bias of 200~V due to the radiation induced creation of extra generation centres. The capacitance measurements showed a dependence upon the test signal frequency which is a characteristic of deep levels. The capacitance measured at 5~V reverse bias with a test frequency of 100~Hz fell with radiation from 300~pF to 40~pF due to the removal of measurable free carriers. The charge collection of the device determined from front alpha illumination also fell to $32\pm5$\% at a reverse bias of 200~V.

\end{abstract}
\end{titlepage}

\section{Introduction}

The possible use of GaAs detectors in the future high energy physics experiment ATLAS\cite{atlas} at the LHC imposes severe demands on the radiation hardness of the detectors.
Semi-insulating undoped GaAs detectors have shown significant charge collection degradation after a 24~GeV/c proton fluence of order $1\times10^{14}$~cm$^{-2}$\cite{oshea,marcus}. The increase in the reverse current after this fluence, however, is slight, (in the region of 2-3 times greater). This charge collection reduction can not be tolerated and requires data on the radiation hardness of other GaAs materials. Epitaxial material with a low initial deep level concentration, and thus high pre-irradiation charge collection efficiency (cce), could provide the necessary radiation tolerance and therefore this material was investigated.

GaAs detectors are also ideally suited as X-ray imaging detectors for medical applications due to their high atomic number. To obtained the best X-ray images a high charge collection efficiency is required, preferably at low bias, and thus a low concentration of deep levels is desired. Epitaxial GaAs should be ideal for this application and this application contributed to the motivation of this work.

\section{Experimental}

The low pressure vapour phase epitaxial GaAs was grown by Epitronics Corporation\cite{epitronics} on the surface of a 450~$\mu$m thick Hitachi Cable Horizontal Bridgeman GaAs substrate. The substrate was n$^{+}$ in nature with silicon doping at a concentration of 10$^{18}$~cm$^{-3}$. The VPE layer was n-type and varied in thickness across the wafer from approximately 240~$\mu$m to 125~$\mu$m. The surface of the VPE layer was chemo-mechanically polished with a H$_{2}$O$_{2}$:NH$_{4}$ solution at a pH of $7.9\pm0.1$, a process used in the preparation of LEC wafers at Glasgow University\cite{mythesis}. This process removed approximately 30~$\mu$m from the VPE layer and reduced surface imperfections. 

Circular diodes of 4~mm is diameter were fabricated with metallic electrodes evaporated onto the front (VPE side) and back of the material. The front Schottky contact was a Ti/Pd/Au multi-layer while the back metallisation was Pd/Ge. The rear contact was not annealed and the ohmic contact was realised due to the n$^{+}$ substrate.

Electrical and charge collection characterisation were performed on the diodes. The electrical characterisations were carried out at 300~K with the use of a Keithley-237 source measurement unit and a Hewlett Packard 4274A multi-frequency LCR meter. The response of the diodes to alpha particle irradiation from an Am-241 source under vacuum and to 60~keV photons from an encapsulated Am-241 source were determined. Since the front of the alpha source was coated with surface contamination, which reduced the energy of the alpha particles, the average energy was determined with an over-depleted silicon detector to be 4.1~MeV. The diodes were biased and read-out with an EG$\&$G Ortec-142 pre-amplifier. The signal was shaped with an Ortec-485 post-amplifier with a shaping time of 500~ns and sent to a multichannel analyser. 

Four of the VPE diodes were exposed to a fluence of $1.25\times10^{14}$ 24~GeV/c protons~cm$^{-2}$ to test their radiation hardness. The irradiation was performed at the CERN PS\cite{ps}, Switzerland, at room temperature and without bias. The diodes were stored for three months at room temperature after irradiation before characterisation. No search was made for effects due to annealing over this period.

\section{Pre-irradiation Characteristics}

\begin{figure}
\begin{center}
\begin{tabular}{cc}
\subfigure[Low bias]%
{\resizebox{.45\textwidth}{!}{\includegraphics{{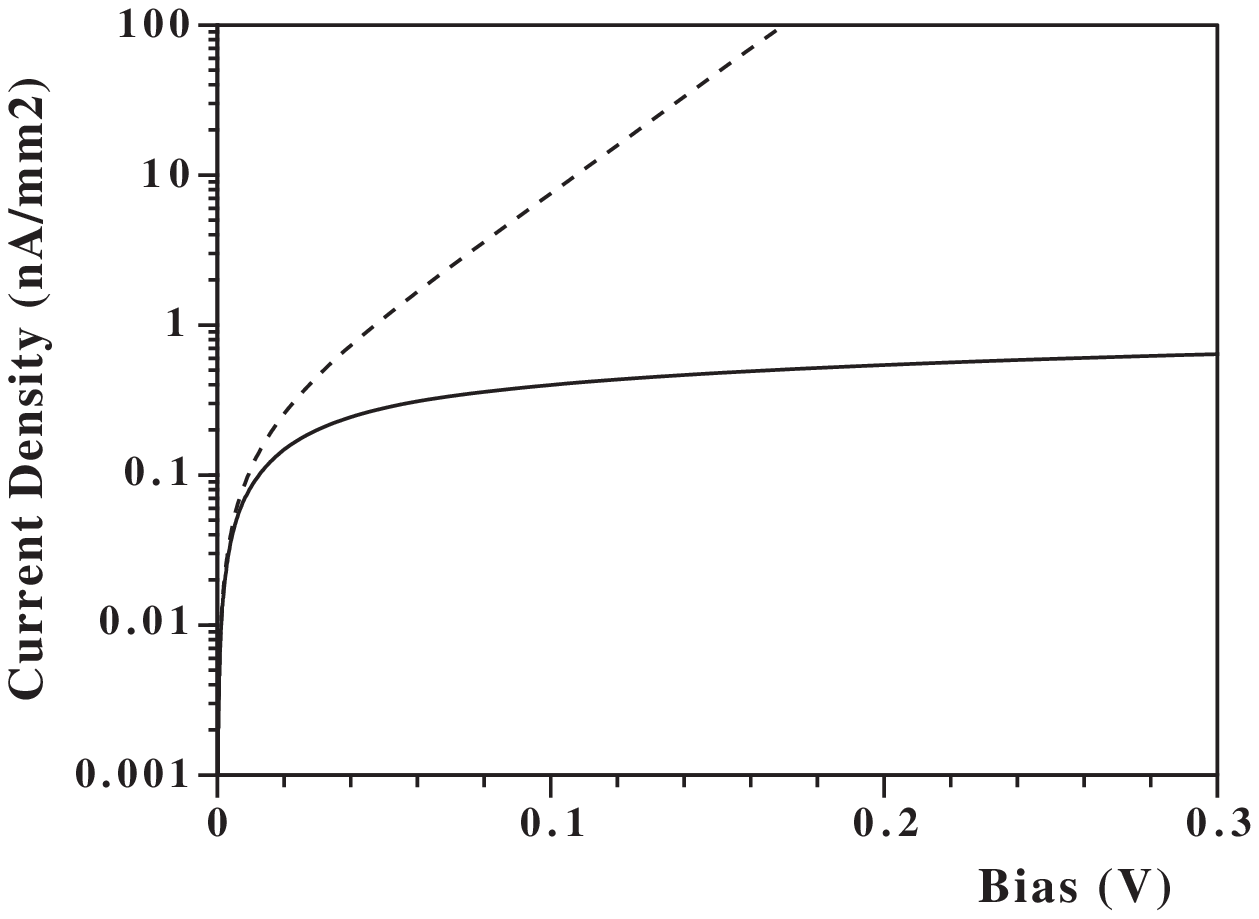}}}\label{fig:pre-iv-lowV}}
\subfigure[High bias]%
{\resizebox{.45\textwidth}{!}{\includegraphics{{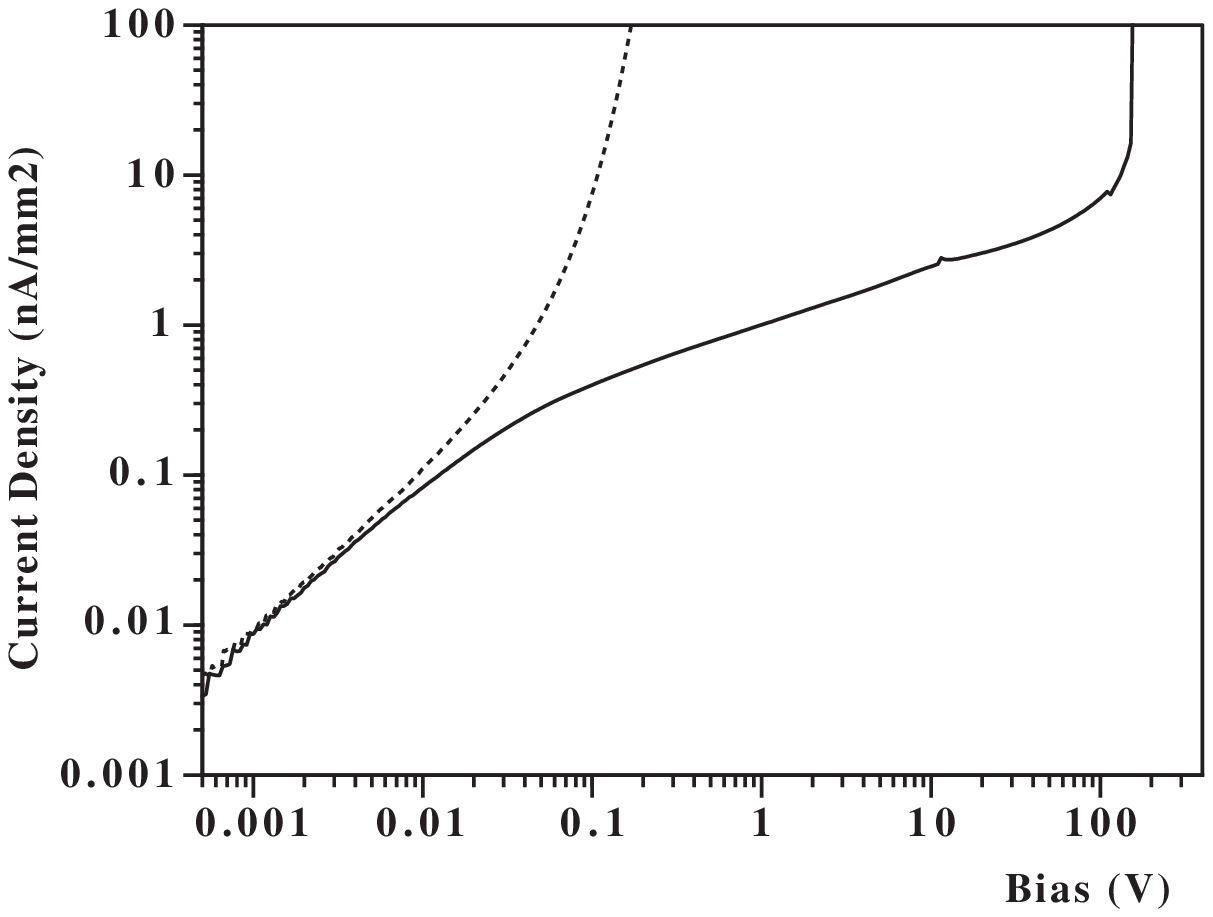}}}\label{fig:pre-iv-highV}}
\end{tabular}
\end{center}
\caption{The current characteristics of a VPE GaAs diode measured at 300~K. The key is: dashed line - forward bias; solid line - reverse bias.}
\label{fig:pre-iv}
\end{figure}

Figure \ref{fig:pre-iv} shows the current-voltage characteristics of the VPE Schottky diode under both forward and reverse bias and demonstrate standard Schottky characteristics for low bias, as shown in figure \ref{fig:pre-iv-lowV}. However, at higher bias the reverse current did not saturate and continued to increase until diode breakdown at approximately 180~V. 

\begin{figure}
\centering
\resizebox{.45\textwidth}{!}{\includegraphics{{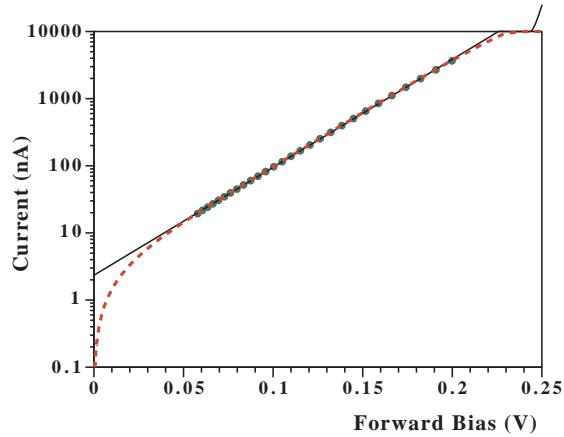}}}
\caption{The extrapolation of the forward current to determine $J_{S}$.}
\label{fig:pre-phib1}
\end{figure}

From the forward bias characteristic two methods were used to determine the barrier height\cite{sze}. The first found the value of the saturation current ($J_{s}$) from the extrapolation of the current characteristics to zero forward bias, shown in figure \ref{fig:pre-phib1}. Using equation \ref{equ:js} with a Richardson constant ($A^{**}$) of 8~Acm$^{-2}$K$^{-2}$, the barrier height ($\phi_{B}$) was found to be $0.81\pm0.005$~V, where the error was due to the uncertainty of $A^{**}$.
\begin{equation}
J_{s} = A^{**}T^{2} \exp\left( - \frac{q \phi_{B}}{kT}\right) 
\label{equ:js}
\end{equation}
This method requires the exact dimensions of the contact area ($S_{e}$) to determine the current density. The second method, which utilises the temperature dependence of the forward current, has no such dependence. The current was measured at 0.1~V over the temperature range 293 to 308~K. The forward current may be expressed as:
\begin{equation}
\ln \left(\frac{I}{T^{2}} \right) = \ln {A^{**} S_{e}} - \frac{q}{k T} (\phi_{b} - V)
\label{equ:iforward}
\end{equation}
The barrier height for the VPE diode was found from a plot of $\ln (I/T^{2})$ against $1/T$ to equal $0.85\pm0.01$~V. The error was found from the error in the linear regression fit to the data.

Capacitance measurements were performed on the diodes. The C-V curves did not show a significant difference between the measurements made with a test signal frequency of 100~Hz and 100~kHz, therefore the concentration of deep levels with a low emission frequency was assumed to be negligible. A doping density ($N_{D}$) of $2.8\pm0.2\times10^{14}$~cm$^{-3}$ for the VPE layer was determined from the capacitance voltage characteristics of the device over a reverse bias range 0-20~V. Such a high value has implications for the extension of the depletion width of the device under reverse bias and thus the charge collection of the device.

\begin{figure}
\centering
\resizebox{.6\textwidth}{!}{\includegraphics{{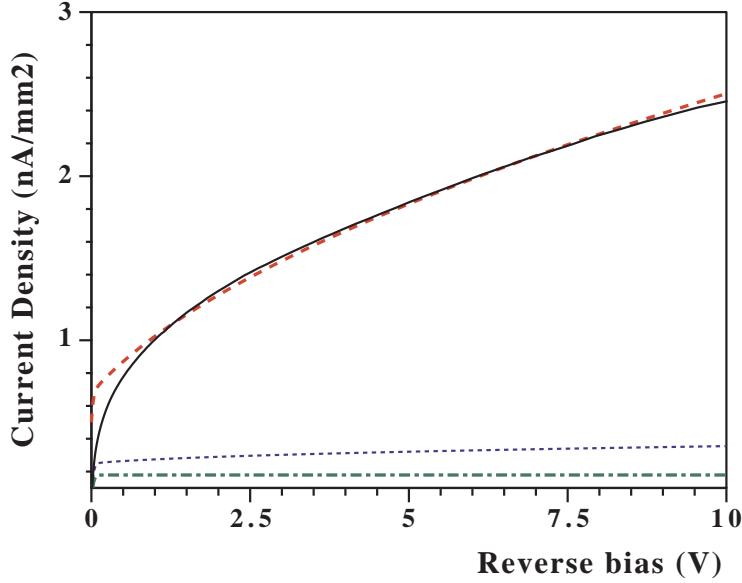}}}
\caption{The reverse bias current-voltage characteristic. The key is: solid line - measured data; dot-dashed line - ideal current; dotted line - that due to image force lowering; dashed line - the result of the fit of equation (\protect\ref{equ:j-gen}).}
\label{fig:pre-iv-10V}
\end{figure}

The reverse current characteristic was investigated to determine the cause of non-saturation. Figure \ref{fig:pre-iv-10V} shows the current voltage dependence up to a reverse bias of 10~V. Illustrated on the figure is the current that would be expected for an ideal barrier. As can be seen this is considerably less than that measured. The effect on the current due to image force lowering of the barrier height for the measured doping density of the material is shown as the dotted line in the figure. This was insufficient to account for the observed current. The current was assumed to be generation current and a fit of the form:
\begin{equation}
J_{gen} = \frac{q n_{i} w}{2 \tau_{r}} = P(1) (\phi_{b} - V_{n} - kT/q + V_{r})^{P(2)}
\label{equ:j-gen}
\end{equation}
was performed to the data, where $P(1)$ and $P(2)$ were free parameters, $n_{i}$ the intrinsic carrier concentration, $w$ the depletion width, and $\tau_{r}$ the lifetime of the free carriers. The value of $V_{n}$ was calculated from:
\begin{equation}
V_{n} = \frac{kT}{q } \ln \left( \frac{N_{C}}{N_{D}}\right)
\label{equ:V_n}
\end{equation}
where $N_{C}$ is the effective density of states in the conduction band.

The values obtained for the parameters were: $P(1) = 0.66\pm0.02$~nA/mm$^{2}$ and $P(2) = 0.53\pm0.02$. For generation current the expected value of $P(2)=0.5$ which corresponds to that found and therefore the current was attributed to generation effects. The activation energy of the current at -10V was found to equal $0.73\pm0.02$~eV which is equal, within errors, to half the bandgap energy. The generation current was not likely to be due to the mid-gap donor EL2, as is the case in LEC material, because the concentration of this deep level was low.

\begin{figure}
\centering
\resizebox{.45\textwidth}{!}{\includegraphics{{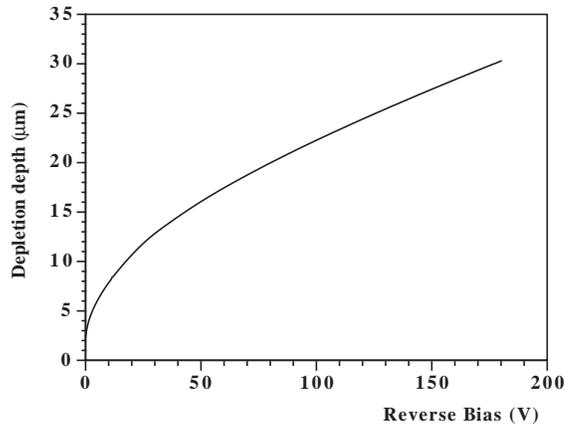}}}
\caption{The depletion width as a function of reverse bias, calculated from the capacitance voltage dependence of the device.}
\label{fig:pre-w}
\end{figure}
\begin{figure}
\begin{center}
\begin{tabular}{cc}
\subfigure[Function of reverse bias]%
{\resizebox{.45\textwidth}{!}{\includegraphics{{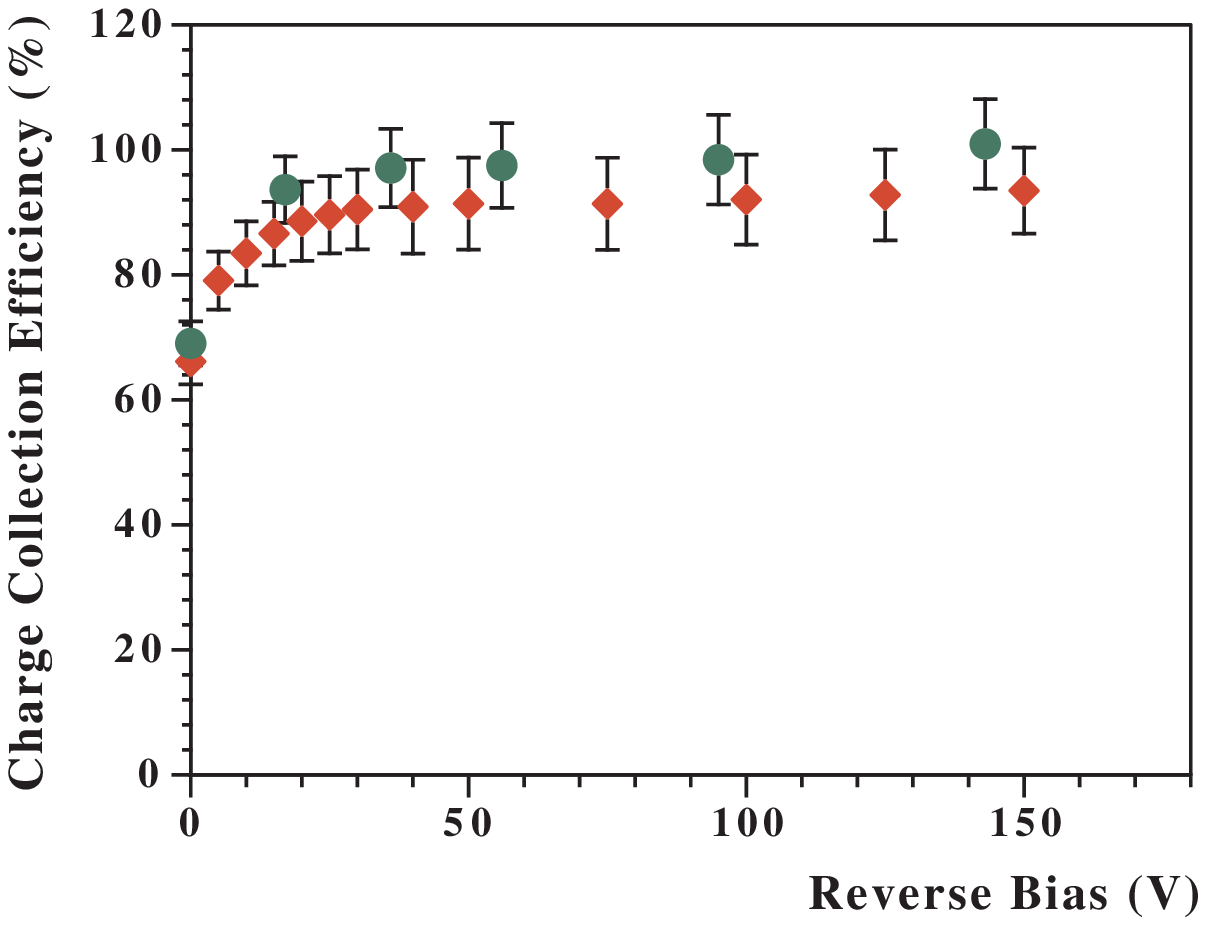}}}\label{fig:pre-alphacceV}}
\subfigure[Function of depletion width]%
{\resizebox{.45\textwidth}{!}{\includegraphics{{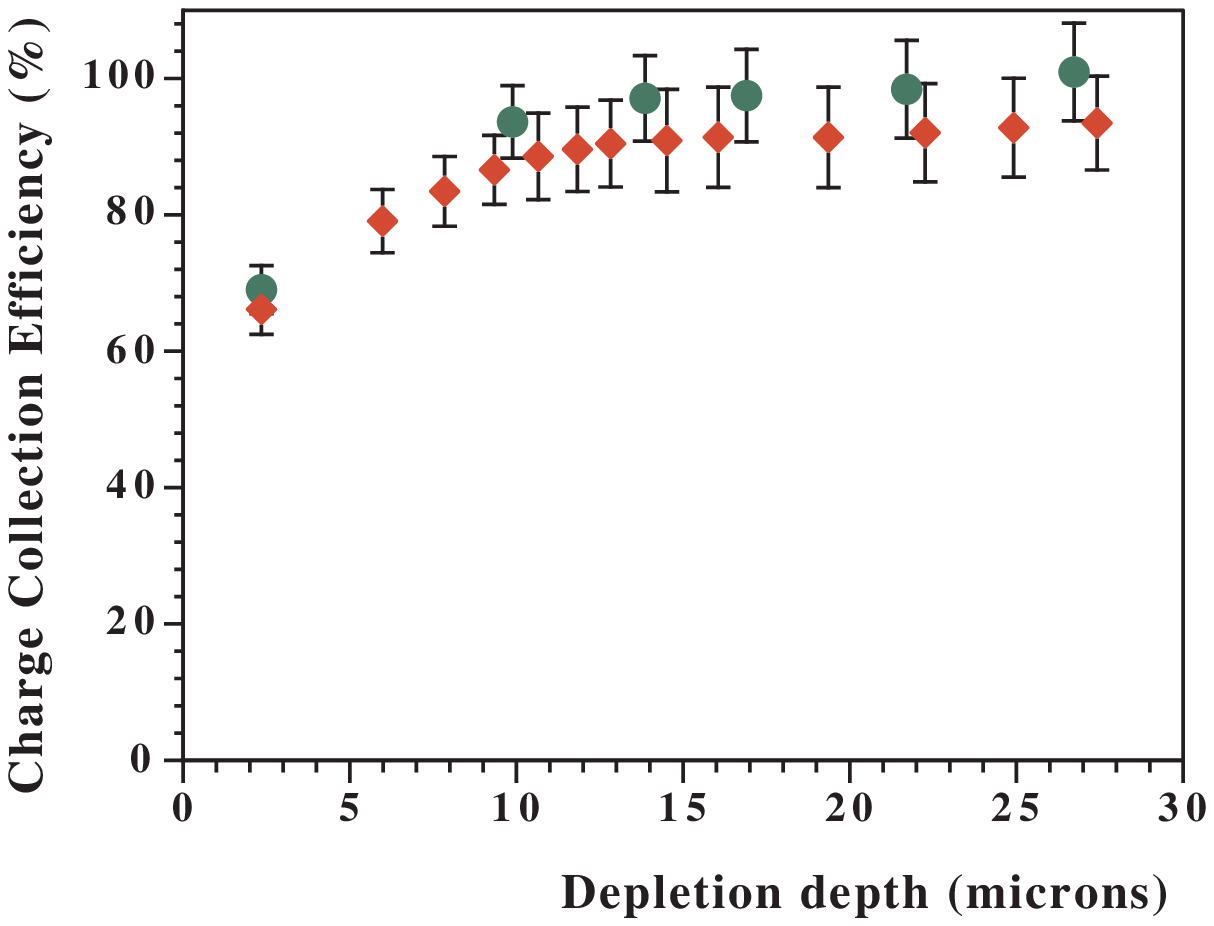}}}\label{fig:pre-alphaccew}}
\end{tabular}
\end{center}
\caption{The charge collection efficiency from front illumination by 4.1~MeV alpha particles as a function of reverse bias and depletion depth for two VPE diodes.}  
\label{fig:pre-alphacce}
\end{figure}

From the capacitance voltage measurements the depletion width as a function of bias was determined up to a bias of 180V, as shown in figure \ref{fig:pre-w}. At 100~V the depletion depth was only 22.3~$\mu$m, increasing to 27.4~$\mu$m at 150~V. The low values were due to the high density of donors in the material.
 The low depletion depth and the high current resulted in the signal from a high energy beta particle being indistinguishable from the electronic noise of the device. For a 100\% cce the signal would equal only 3600 electrons at a bias of 150~V. The cce determined for 60~keV gamma photons was $97\pm17$\% at a bias of 200V. The error is due to the spread in the peak due to the high leakage current of the device at this bias. Due to the 450~$\mu$m thick $n^{+}$ layer on the back side of the device the detector was only sensitive to alpha particle illumination from the front surface. Although the range of 4.1~MeV alpha particles in GaAs is only $\sim13\mu$m, due to the comparably small depletion region the detector response was no longer due primarily to one charge carrier. The charge collection efficiency as a function of both bias and depletion depth, calculated from the CV data, are shown in figure \ref{fig:pre-alphacce} for two VPE diodes. The charge collection showed the expected increase with bias due to the increase in depletion width. The maximum cce's for the two detectors were $100\pm$7\% and $93\pm$7\% at a bias close to 150~V. The cce plateaued for a depletion width between 10~$\mu$m and 15~$\mu$m as expected from the alpha particle range in GaAs. At zero applied bias, however, the depletion width, calculated from the capacitance measurements, was only 2~$\mu$m but a signal of 65\% of the full alpha particle signal was obtained. Therefore diffusion or drift of charge into the depletion region must occur.

\section{Post-irradiation Characteristics}

\begin{figure}
\centering
\resizebox{.45\textwidth}{!}{\includegraphics{{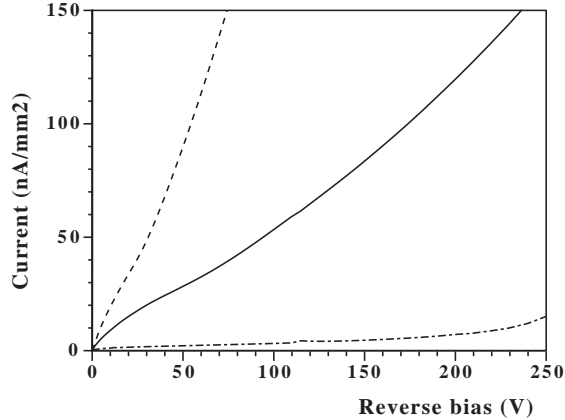}}}
\caption{The current-voltage characteristics of the VPE diodes before and after a 24GeV/c proton fluence of $1.25\times10^{14}$~cm$^{-2}$, measured at 300~K. The key is: dot-dashed line - reverse current before; solid line - reverse current after; dashed line - forward current after irradiation. The forward current before irradiation is indistinguishable from the current axis on this figure.}
\label{fig:post-iv}
\end{figure}

Figure \ref{fig:post-iv} illustrates the significant change in the current characteristics after the $1.25\times10^{14}$~cm$^{-2}$ proton fluence. The reverse current increased in the operating region of the detector and was $\approx$85~nA/mm$^{2}$ at a reverse bias of 150~V compared to only 4.6~nA/mm$^{2}$ before irradiation. The forward current on the other hand was reduced after the fluence. The current characteristic was more resistive in nature and thus the resistivity of the material had increased. This implies that the product of the mobility and the free carrier concentration was reduced. A reduction in mobility is very likely as the increased number of defect centres will increase the concentration of scattering centres and thus reduce the mobility. The free carrier concentration will be reduced if deep levels are introduced which compensate the material and thus cause the Fermi level to approach mid-bandgap. The activation energy of the reverse current was measured at a bias of -10V and -100V to be $0.57\pm0.01$~eV if generation current was assumed. The change in activation energy was due to the presence of radiation induced deep levels.

\begin{figure}
\begin{center}
\begin{tabular}{cc}
\subfigure[After irradiation performed with two test signal frequencies. The key is: dotted line - 100~kHz; solid line - 100~Hz.]%
{\resizebox{.45\textwidth}{!}{\includegraphics{{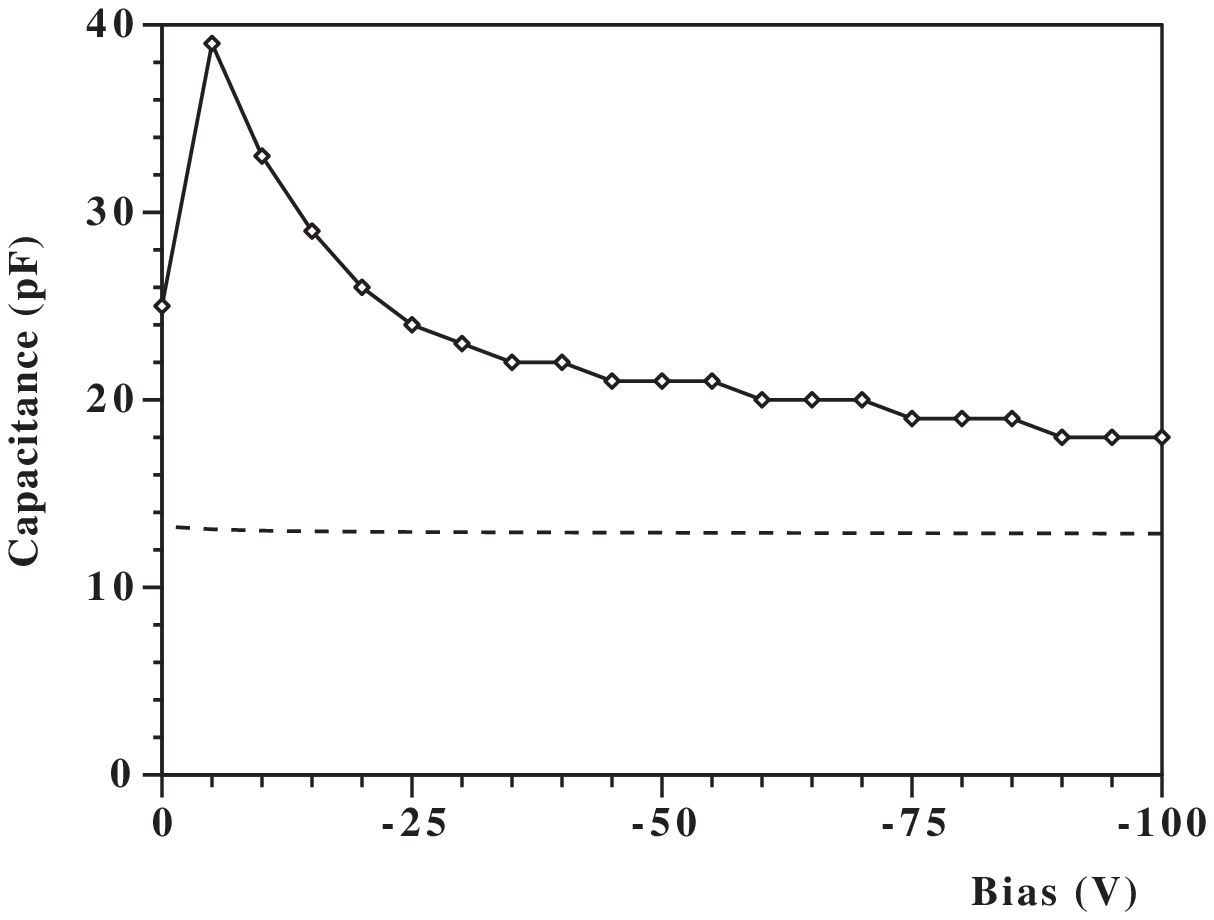}}}\label{fig:post-cv}}
\subfigure[Before and after irradiation measured with a test signal of 100~Hz. The key is: solid line - pre; dotted line - post.]%
{\resizebox{.45\textwidth}{!}{\includegraphics{{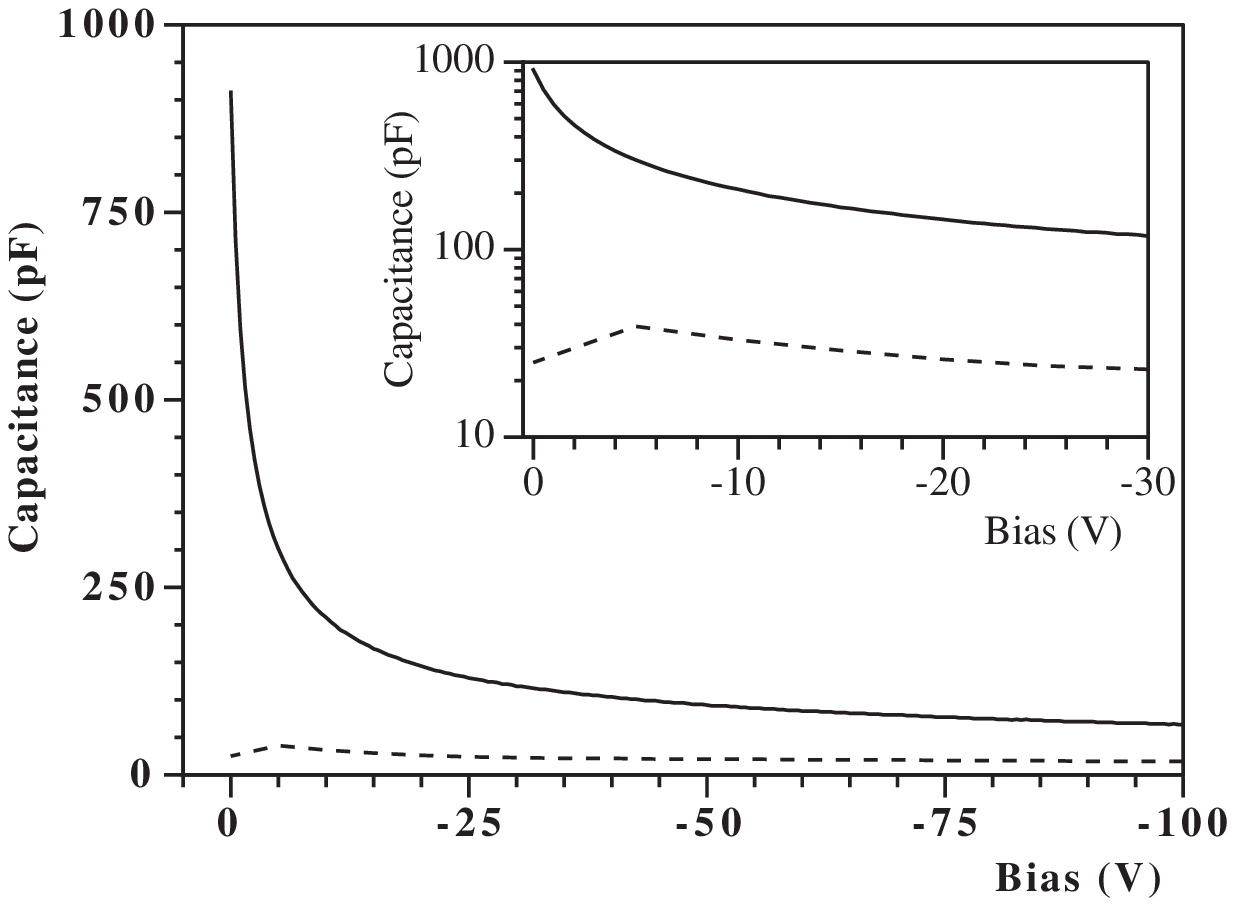}}}\label{fig:post-cv2}}
\end{tabular}
\end{center}
\caption{Capacitance-voltage measurements before and after irradiation, measured at 20$^{o}$C.}
\end{figure}

The capacitance-voltage relationship shows a dependence upon the frequency of the test signal used to make the measurement, illustrated in figure \ref{fig:post-cv}. The high frequency capacitance did not change with bias. This is characteristic of measurements made on high resistivity material where the capacitance measured with a high frequency test signal equals the geometrical capacitance of the device\cite{mythesis}. For this diode the measured capacitance at 100~kHz was 13~pF. The thickness of the VPE layer was not measured directly but a device of this capacitance implies that the diode is 110$\mu$m thick, which is in the expected range for the thickness of the VPE material.

\begin{figure}
\centering
\resizebox{.45\textwidth}{!}{\includegraphics{{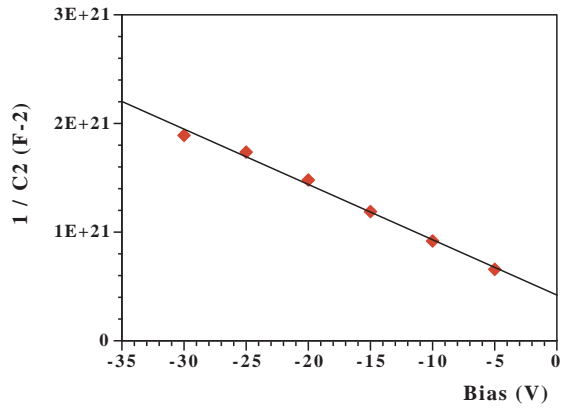}}}
\caption{1/C$^{2}$ as a function of bias for an irradiated VPE diode.}
\label{fig:post-1/c2}
\end{figure}

The capacitance voltage dependence before and after irradiation are compared in figure \ref{fig:post-cv2}. The measured capacitance at a given bias fell and at -5~V was only 40~pF compared to 300~pF. As the measured capacitance depends upon the density of free carriers and the density of traps which have emission rates greater than the frequency of the A.C. test voltage, the reduction in the measured capacitance implies that the density of these carriers was reduced. Therefore a significant concentration of deep levels was created in the material due to the proton fluence. The effective trap concentration ($N_{eff}$) was deduced from the 1/C$^{2}$ bias voltage dependence shown in figure \ref{fig:post-1/c2}, measured between -5~V and -30~V, to be only $1.3\times10^{13}$~$cm^{3}$. The reduction in free carrier density measured as a fall in $N_{eff}$ and as an increase in resistivity implies that either acceptor (electron) traps have been introduced and/or that the Fermi level has been lowered, resulting in a lower free carrier concentration. Lowering of the Fermi level will be caused by an increased concentration of deep levels that remove electrons from the conduction band, that is acceptor levels.   


\begin{figure}
\centering
\resizebox{.45\textwidth}{!}{\includegraphics{{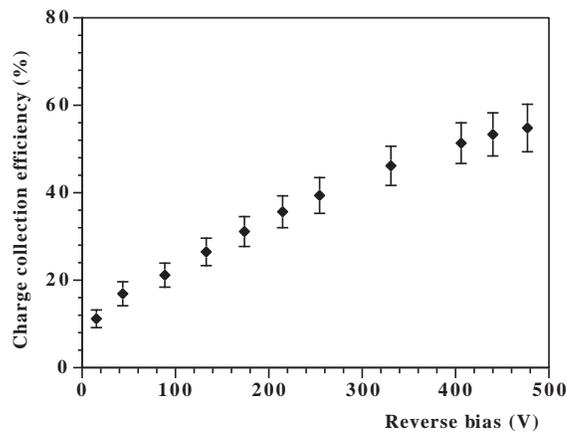}}}
\caption{Charge collection efficiency from front alpha illumination after irradiation.}
\label{fig:post-alphacce}
\end{figure}

The charge collection efficiency of the irradiated device was investigated. Again results from high energy beta particles could not be obtained due to leakage current noise. The increase in reverse current also prevented the detection of 60~keV photons. Charge collection from front illumination by alpha particles was obtained and is illustrated in figure \ref{fig:post-alphacce} as a function of bias. The cce fell from $\sim$100\% before irradiation to $32\pm4$\% measured at 200V. The diode did not breakdown until higher voltages and a cce of 55\% was measured at the highest possible voltage of 480~V.

\section{Comparison to LEC diodes}

\begin{figure}
\begin{center}
\begin{tabular}{cc}
\subfigure[Before radiation]%
{\resizebox{.45\textwidth}{!}{\includegraphics{{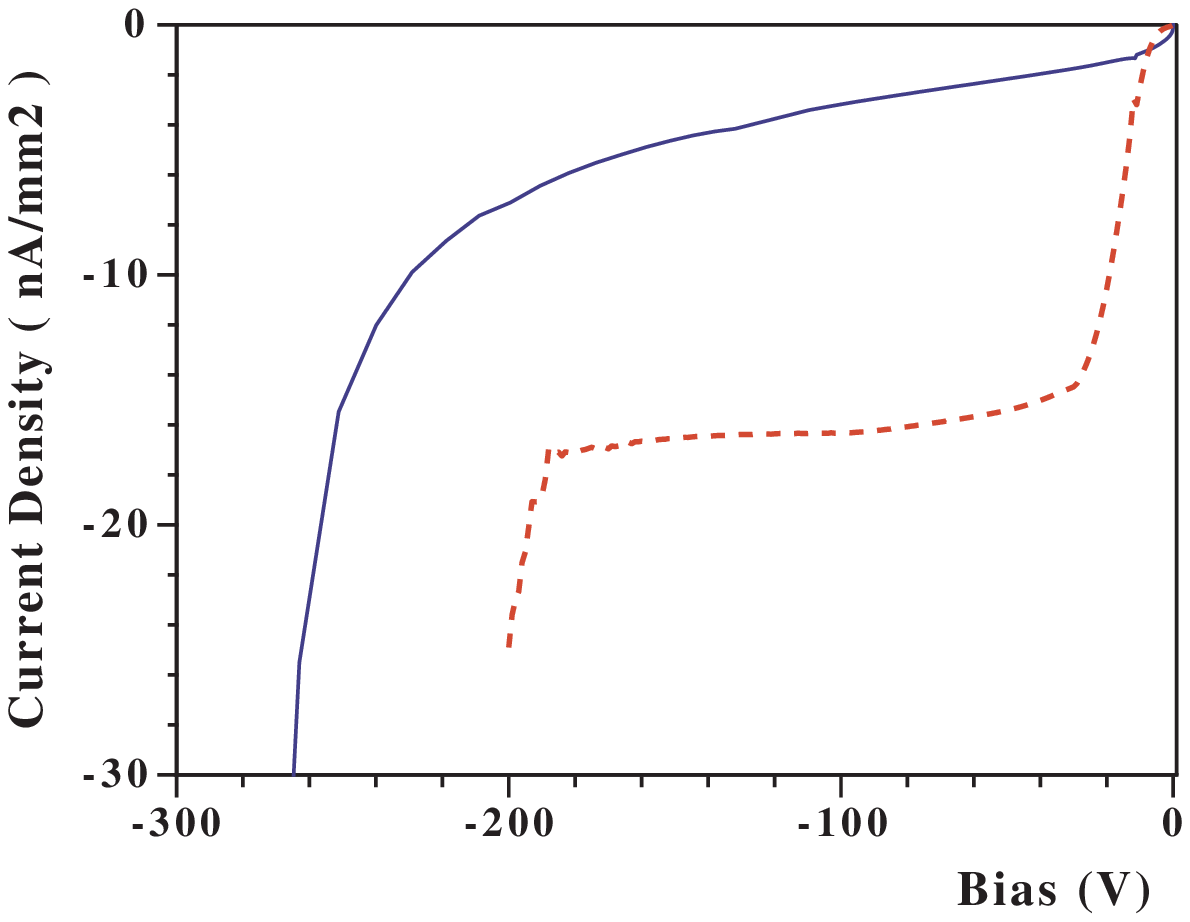}}}\label{fig:comp-iv-pre}}
\subfigure[After radiation]%
{\resizebox{.45\textwidth}{!}{\includegraphics{{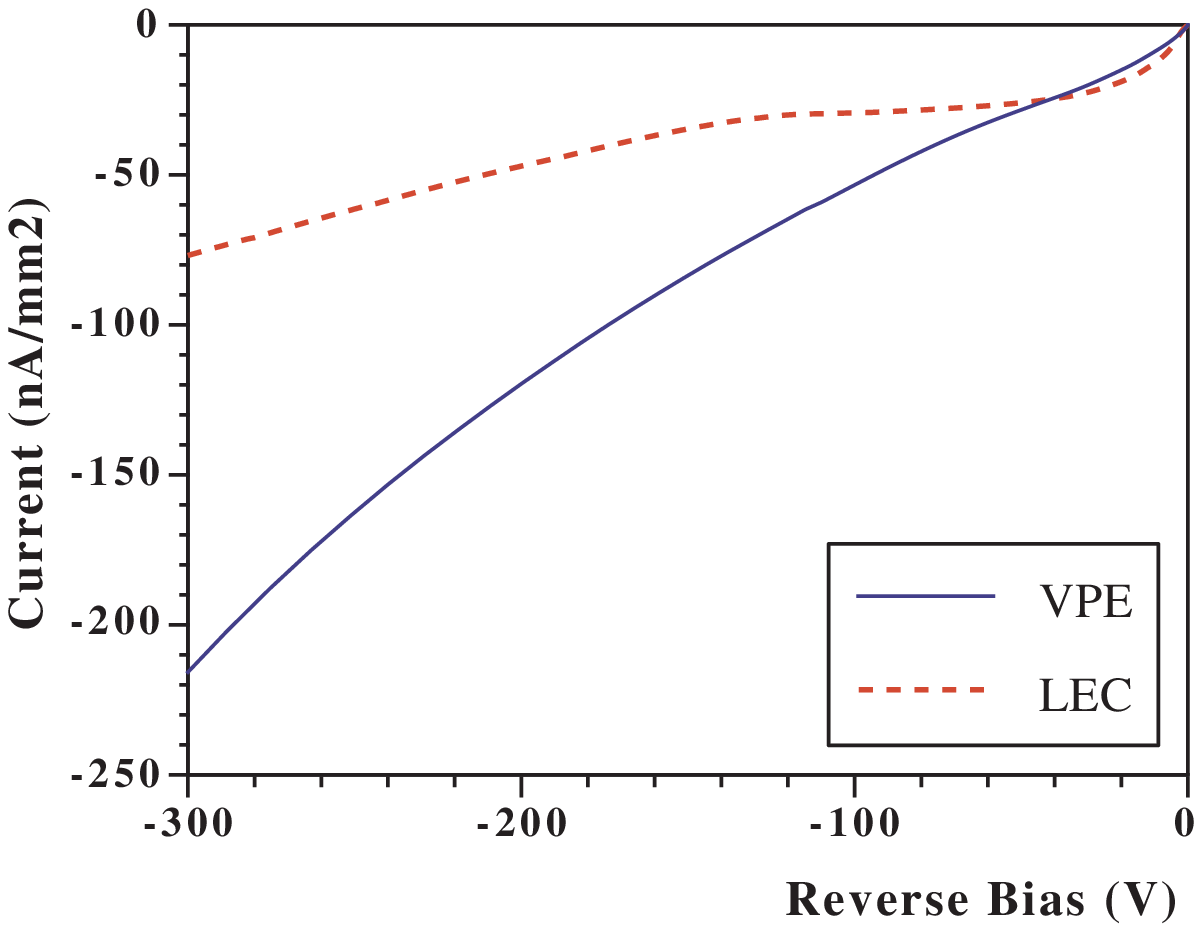}}}\label{fig:comp-iv-post}}
\end{tabular}
\end{center}
\caption{Comparison of the reverse current-voltage characteristics of a VPE and an LEC diode before and after irradiation. The key is: solid line - VPE; dashed line - LEC}
\label{fig:comp-iv}
\end{figure}

The reverse current-voltage characteristics of a VPE diode and an LEC diode before and after the radiation fluence are shown in figure \ref{fig:comp-iv}. Before irradiation the current density is less for the VPE diode than for the LEC diode. After irradiation, however, the reverse current of the VPE diode increased dramatically by over an order of magnitude, but, for the LEC diode the increase was only by a factor of two at a reverse bias of 100~V.

The charge collection of the VPE diode was higher than the LEC diode both before and after irradiation. It should be noted, however, that the VPE diode had a depletion width of only $\sim$30$\mu$m and therefore the effect of a short absorption length will be less significant.

\section{Conclusions}

The Schottky diodes fabricated with VPE GaAs behaved according to simple theory, that is without the presence of a significant concentration of deep traps. Titanium-GaAs barrier heights of $0.81\pm0.005$eV and $0.85\pm0.01$~eV were measured which are closed to those quoted in the literature. The reverse current-voltage characteristic did not saturate, an effect attributed to generation current in the depletion region. The donor concentration was obtained via CV analysis of the material to be $2.8\pm0.2\times10^{14}$~cm$^{-3}$, which prevented the extension of the depletion region beyond $\sim$30~$\mu$m before breakdown occurred, due to the high bias required ($>$200~V). Charge collection experiments were limited due to the small depletion region. The charge collection efficiencies obtained from front alpha particle and 60keV gamma irradiation were close to 100\%.

After a 24GeV/c proton fluence of $1.25\times10^{14}$~cm$^{-2}$ the VPE diode had a reverse current density measured at 100V reverse bias and 300~K, of $5.3\times10^{-6}$~Acm$^{-2}$. This may be compared to $2.8\times10^{-6}$~Acm$^{-2}$ for an LEC diode after the same fluence. The current increase for the VPE diode was over an order of magnitude. It must be noted, however, that the VPE diode was not guarded and a guarded ring could reduce the current due to a reduction in surface currents. Deep levels were introduced. From the increase in resistance and the capacitance-voltage dependence the deep levels were appear to be acceptors. A fall in charge collection efficiency from front alpha illumination was observed and at a reverse bias of 200~V the cce fell from 100\% to $32\pm5$\%.

In comparison with a typical LEC diode the VPE diode had a lower leakage current before irradiation, but this was not true after the radiation fluence. Charge collection from a SI-U detector was considerably worse than in the VPE device both before and after the radiation fluence. The small depletion width of the VPE diode, however, means that the results can not be directly compared.

The material could be very promising for detector development if the free carrier concentration can be reduced to allow an increased depletion width. The materials radiation hardness is still a cause for concern.

\section{Acknowledgements}

The authors would like to thank F.McDevitt, A.Meikle and F.Doherty for technical support and all those at the CERN PS facility during the irradiation run. One of us (R.Bates) gratefully acknowledges the support received through a CASE postgraduate studentship from RAL. The results obtained within the RD8 collaboration are from work partly funded by PPARC (UK), INFN (Italy), and the BMFD (Germany).

\end{document}